# Prediction modelling with many correlated and zero-inflated predictors: assessing a nonnegative garrote approach

Mariella Gregorich[1], Michael Kammer[1,2], Harald Mischak[3,4] and Georg Heinze[1,*]

[1]Medical University of Vienna, Center for Medical Data Science, Institute of Clinical Biometrics, Vienna, Austria

[2]Medical University of Vienna, Department of Medicine III, Division of Nephrology, Vienna, Austria

[3]University of Glasgow, School of Cardiovascular and Metabolic Health, Glasgow, UK

[4]Mosaiques Diagnostics GmbH, Hannover, Germany

[*]Corresponding author

## Abstract

Building prediction models from mass-spectrometry data is challenging due to the abundance of correlated features with varying degrees of zero-inflation, leading to a common interest in reducing the features to a concise predictor set with good predictive performance given the experiments' resource-intensive nature. In this study, we formally established and examined regularized regression approaches, designed to address zero-inflated and correlated predictors. In particular, we describe a novel two-stage regularized regression approach (ridge-garrote) explicitly modelling zero-inflated predictors using two component variables, comprising a ridge estimator in the first stage and subsequently applying a nonnegative garrote estimator in the second stage. We contrasted ridge-garrote with one-stage methods (ridge, lasso) and other two-stage regularized regression approaches (lasso-ridge, ridge-lasso) for zero-inflated predictors. We assessed the predictive performance and predictor selection properties of these methods in a comparative simulation study and a real-data case study with the aim to predict kidney function using peptidomic features derived from mass-spectrometry. In the simulation study, the predictive performance of all assessed approaches was comparable, yet the ridge-garrote approach consistently selected more parsimonious models compared to its competitors in most scenarios. While lasso-ridge achieved higher predictive accuracy than its competitors, it exhibited high variability in the number of selected predictors. Ridge-lasso exhibited slightly superior predictive accuracy than ridge-garrote but at the expense of selecting more noise predictors. Overall, ridge emerged as a favourable option when variable selection is not a primary concern, while ridge-garrote demonstrated notable practical utility in selecting a parsimonious set of predictors, with only minimal compromise in predictive accuracy.

# 1. Introduction

Prediction modelling involves the identification of predictors that can accurately prognosticate future health outcomes or assist in diagnosing a patient's condition. To that end, modern technologies facilitate the generation of a large number of such predictive factors, for example peptide intensities measured via mass-spectrometry. However, usually a considerable proportion of the collected variables may not be needed for accurate prediction of the particular outcome of interest. Not only the large number of peptides poses a challenge, but in addition the measured intensities from mass-spectrometers are often subject to zero-inflation, i.e., their distribution exhibits a spike at zero, resulting in peptide distributions that need special consideration when used as predictors in a (generalized) linear model. The zero intensities have been termed point mass values (PMVs), while the continuous, nonnegative intensities that are not zero have been referred to as non-PMVs [1]. The reasons for measuring zero values are twofold: (1) structural zeros due to biological absence of a peptide in a sample or (2) sampling zeros due to under-sampling of low-abundance peptides (e.g. when the true intensity is lower than a technical limit of detection, LOD), technical issues (e.g. loss during preparation due to absorbance or signal suppression) or error in the interpretation of raw mass spectral data. Unfortunately, it is impossible to distinguish between these cases objectively in practical settings as the exact biological mechanisms behind the production of peptides are usually the target of investigation and thus often unknown. Ad-hoc techniques dealing with the excess of zeros generally either assume all PMVs to be structural zeros, or treat all PMVs as sampling zeros and substitute them with $LOD\ /\ c$, $LOD/\sqrt{2}$ or $LOD-c$, where $c$ is an arbitrary constant value [2-4]. Nonetheless, both methods fail to address the dual nature of the intensity distribution, leading to potentially biased parameter estimates and inadequate model fit.

Previous research has addressed the challenges associated with the analysis of high-dimensional, zero-inflated and correlated predictors by employing regularized regression models combined with different data pre-processing strategies to handle the excess zeros in the data. For instance, Gajjala et al. [5] used the least absolute shrinkage and selection operator (lasso) to identify proteomic markers of chronic kidney disease (CKD) progression from high-dimensional plasma peptides. The peptide intensity values were subjected to pre-processing, involving a $\log_2$-transformation and accounting for PMVs by subtracting a constant value from the minimum measured values of individual peptides. Further, Pena et al. [6] implemented a two-stage modelling approach to identify plasma proteomics markers. Initially, lasso was used to select differentially expressed peptides between subjects with CKD and those without. Following this, ridge regression was conducted, using two variables for each selected peptide: a binary variable that distinguished between PMVs and non-PMVs, and a continuous variable, representing the logarithm of the non-PMVs. A comparable procedure to derive a proteomic score able to differentiate between patients and controls was carried out in the study of Gajjala et al. [7]. In many cases, algorithmic classifiers such as support vector machines [8, 9] or random forest [10] were employed for diagnostic purposes to sidestep the need for data pre-processing considerations. Nonetheless, the optimal strategy for handling and analysing zero-inflated and correlated predictors is uncertain, particularly in high-dimensional settings.

As mass-spectrometry experiments are costly and time-consuming, it is crucial to identify a technique that offers good predictive performance while identifying a parsimonious set of peptides relevant for prediction. Here, we aimed to explicitly define the methodological background of regularized regression techniques tailored for variable selection or shrinkage in the presence of zero-inflation and a complex correlation structure. Additionally, we examined their capabilities in predictive performance and assessed their predictor selection properties. We considered two-stage regularized regression approaches combining $L_1$- and $L_2$ penalties with the objective to obtain a parsimonious set of predictive peptides. A particular emphasis is put on the evaluation of the nonnegative garrote with ridge estimates in the first stage instead of the standard ordinary least squares estimates (ridge-garrote) [11]. In a case study, we demonstrate the performance of the approaches to predict estimated glomerular filtration rate (eGFR) using mass-spectrometry data.

The following section outlines the methodological background of the regularized regression techniques tailored to address the challenges of predictors with zero-inflation. Following that, an extensive simulation study evaluates their predictive performance and their ability to produce a parsimonious set of predictors in Section 3. Additionally, a case study demonstrates their use in predicting kidney function from proteomic mass-spectrometry data in Section 4. In the last section, we discuss the study's findings, limitations, and draw conclusions.

## 2. Methods

### 2.1. Strategies for handling zero-inflated predictors

Zero-inflated predictors need special treatment in statistical models if one has reason to assume that the expected value of the outcome at a zero value of a predictor cannot be extrapolated from the association of non-zero values of the predictor with the outcome [12]. While we focus on the strategy that represents each predictor by two component variables [6, 13, 14], a naïve approach based on imputation of the PMVs is also briefly outlined. We assume a study cohort of $n$ subjects. Let $\boldsymbol{y} = (y_1, \ldots, y_n)^T$ be the vector of observed outcomes and $\boldsymbol{Z}$ the $n \times q$ matrix of predictors with observed continuous and nonnegative intensity values $z_{ij}, i = 1, \ldots, n; j = 1, \ldots, q$. Let $\boldsymbol{z_j}$ be the $j$th, potentially zero-inflated predictor in $\boldsymbol{Z}$.

**Strategy A - 'component variables':** The predictor $\boldsymbol{z_j}$ may be represented by two component variables:

I. <u>a binary component</u>: $\boldsymbol{d_j} \coloneqq I(\boldsymbol{z_j} > 0)$, where $I(\cdot)$ denotes the indicator function for the presence or absence of an intensity value higher than 0, and

II.  <u>a continuous component</u>: $u_j := f(z_j)$, where $f(\cdot)$ represents a monotone function that aims to symmetrize the distribution of the intensities (such as the logarithm, or the square root, but may also be the identity function). If $f(0)$ is undefined, PMVs are substituted with a common representative value that can be chosen arbitrarily.

The binary component matrix $D$ addresses the PMVs in the data, thereby allowing to model the zeros and non-zero values separately and accounting for potentially different processes that generated these intensities. The continuous component matrix $U$ comprises the non-PMVs. When employing a modelling approach that can only select $u_j$, the choice of the representative value of the PMVs is crucial to prevent information loss caused by the integration of PMVs into the distribution of the non-PMVs, thereby complicating the interpretation of the corresponding coefficient in the model. Strategy A is useful to handle PMVs if they are a mixture of structural and sampling zeros.

**Strategy B - 'imputation of PMVs':** we may define the matrix $X$, which consists of the normalized values of $Z$ after replacing PMVs with a fixed, biologically meaningful value larger than 0 (e.g. $LOD/2$, $LOD$ or $LOD/\sqrt{2}$). This imputation assumes that all PMVs are sampling zeros. If PMVs are a mixture of structural and sampling zeros, a model that considers only $X$ for the peptide data imposes the strong assumption that structural zeros of the peptides behave exactly like sampling zeros.

The matrix $X$ is obtained by a similar normalization transformation as $U$ and differs from $U$ merely in the value chosen to impute PMVs. While for $U$, imputations for PMVs (e.g. the mean of the normalized non-PMVs) should support interpretability of the regression coefficients of the binary components $d_j$, for $X$ PMVs are replaced by biologically meaningful values. However, the strategies are not mutually exclusive; one may consider to use matrices $X$ and $D$ to represent the candidate variables which would allow to obtain interpretable models even if $X$ and $D$ components are independently selected. We refer to the data example in Section 4 for further details on the data transformation in an applied setting. In the following sections, we assume the transformations have been applied to the observed values of the intensity distribution $Z$, following Strategy A to produce component matrices $U$ and $D$ with all binary and continuous variables, and following Strategy B to generate the transformed matrix $X$.

## 2.2. One-stage regression models

As commonly used in practice, we investigated <u>lasso regression</u> [15] which provides selection and shrinkage of the regression coefficients for the candidate predictors based on L$_1$-regularization. In the case of linear regression, it minimizes the criterion

$$\widehat{\boldsymbol{\beta}}_X^{lasso}(\lambda) = \underset{\boldsymbol{\beta}_X}{\operatorname{argmin}} \|\boldsymbol{y} - \boldsymbol{X}\boldsymbol{\beta}_X\|_2^2 + \lambda\|\boldsymbol{\beta}_X\|_1, \tag{1}$$

where $\boldsymbol{\beta}_X$ are the unknown regression parameters corresponding to $X$, and $\lambda$ is a parameter to control the amount of regularization.

As another common choice in practice, we included a method based on ridge regression, in which the regression parameters are estimated by

$$\widehat{\boldsymbol{\beta}}^{ridge}(\lambda) = \underset{\boldsymbol{\beta}}{\operatorname{argmin}} \|\boldsymbol{y} - (\boldsymbol{U}\boldsymbol{\beta}_U + \boldsymbol{D}\boldsymbol{\beta}_D)\|_2^2 + \lambda(\|\boldsymbol{\beta}_U\|_2^2 + \|\boldsymbol{\beta}_D\|_2^2), \qquad (2)$$

where $\boldsymbol{\beta} := (\boldsymbol{\beta}_U^T, \boldsymbol{\beta}_D^T)^T$ in which $\boldsymbol{\beta}_U$ and $\boldsymbol{\beta}_D$ are the regression parameter vectors for $\boldsymbol{U}$ and $\boldsymbol{D}$, respectively. Whilst lasso is capable of performing predictor selection, unlike ridge regression, it has been noted that ridge regression is better suited to handle correlated predictors [16].

### 2.3. Two-stage regression approaches

In the lasso-ridge approach, the first stage applies the lasso on the candidate variables represented by $\boldsymbol{X}$. At the second stage, the selected predictors $\widetilde{\boldsymbol{X}} \subseteq \boldsymbol{X}$, are subjected to a ridge regression model but are represented in this model by their corresponding components $\widetilde{\boldsymbol{U}} \subseteq \boldsymbol{U}$ and $\widetilde{\boldsymbol{D}} \subseteq \boldsymbol{D}$:

$$\widehat{\boldsymbol{\beta}}^{lasso-ridge}(\lambda) = \underset{\boldsymbol{\beta}}{\operatorname{argmin}} \left\|\boldsymbol{y} - (\widetilde{\boldsymbol{U}}\boldsymbol{\beta}_{\widetilde{U}} + \widetilde{\boldsymbol{D}}\boldsymbol{\beta}_{\widetilde{D}})\right\|_2^2 + \lambda(\|\boldsymbol{\beta}_{\widetilde{U}}\|_2^2 + \|\boldsymbol{\beta}_{\widetilde{D}}\|_2^2), \qquad (3)$$

where $\boldsymbol{\beta} := (\boldsymbol{\beta}_{\widetilde{U}}^T, \boldsymbol{\beta}_{\widetilde{D}}^T)^T$, with the $\tilde{q} \leq q$-dimensional vectors of the selected component variables $\boldsymbol{\beta}_{\widetilde{U}}$ and $\boldsymbol{\beta}_{\widetilde{D}}$.

In the ridge-lasso approach, the first stage fits a ridge regression as in Equation (2). Then, a weight term for each predictor is defined as the reciprocal of the sum of the absolute values of the corresponding estimated coefficients of the component variables from the first-stage ridge model:

$$w_j = \frac{1}{\left(\left|\widehat{\boldsymbol{\beta}}_{U_j}\right| + \left|\widehat{\boldsymbol{\beta}}_{D_j}\right|\right)}, j = 1, \dots, q \qquad (6)$$

where $\widehat{\boldsymbol{\beta}}_{U_j}$ denotes the regression coefficient corresponding to the $j$th component variable in $\boldsymbol{U}$, and $\widehat{\boldsymbol{\beta}}_{D_j}$ represents the regression coefficient for the $j$th component variable in $\boldsymbol{D}$. At the second stage, a lasso fit is obtained utilizing these weights as peptide-specific penalization factors by minimizing

$$\widehat{\boldsymbol{\beta}}^{ridge-lasso}(\lambda) = \underset{\boldsymbol{\beta}}{\operatorname{argmin}} \|\boldsymbol{y} - (\boldsymbol{X}\boldsymbol{\beta}_X + \boldsymbol{D}\boldsymbol{\beta}_D)\|_2^2 + \lambda(\|\boldsymbol{w}\boldsymbol{\beta}_X\|_1 + \|\boldsymbol{w}\boldsymbol{\beta}_D\|_1). \qquad (7)$$

where $\boldsymbol{\beta} := (\boldsymbol{\beta}_X^T, \boldsymbol{\beta}_D^T)^T$. For the ridge-lasso approach, we substitute the component variable $\boldsymbol{U}$ by $\boldsymbol{X}$ at the second stage because the two components are independently selected in (7). If for a peptide the binary component $\boldsymbol{D}$ is not selected the regression coefficient of the remaining $\boldsymbol{X}$ component can still be interpreted.

In the ridge-garrote approach, the first stage fits a ridge regression model with $\boldsymbol{U}$ and $\boldsymbol{D}$ to obtain $\widehat{\boldsymbol{\beta}}_U^{ridge}$ and $\widehat{\boldsymbol{\beta}}_D^{ridge}$ [11, 17]. At the second stage, shrinkage factors for all $q$ predictors are estimated by solving

$$\hat{c}(\lambda) = \underset{c}{\arg\min} \|y - (U\hat{B}_U + D\hat{B}_D)c\|_2^2 + \lambda\|c\|_1, \quad \text{subject to } c_j \geq 0 \text{ and } \lambda \geq 0, \tag{8}$$

where $\hat{B}_U$ and $\hat{B}_D$ are diagonal matrices of the initial regression coefficients $\hat{\beta}_U^{ridge}$ and $\hat{\beta}_D^{ridge}$ for $U$ and $D$, respectively. The ridge-garrote estimates for both components of each predictor can then be computed by

$$\hat{\beta}_U^{ridge-garrote}(\lambda) = \hat{c}(\lambda)\hat{\beta}_U^{ridge} \quad \text{and} \quad \hat{\beta}_D^{ridge-garrote}(\lambda) = \hat{c}(\lambda)\hat{\beta}_D^{ridge}. \tag{9}$$

For a given predictor, the second step assigns a common shrinkage factor to both components. If this shrinkage factor is estimated as 0, the predictor is deselected from the model.

## 3. Simulation study

We used the structured approach of ADEMP (aim, data, estimands, methods, performance) to report the simulation study [18].

### 3.1. Aims

Our study aimed to evaluate the predictive performance and variable selection properties of five regularized regression methods (ridge, lasso, ridge-lasso, lasso-ridge and ridge-garrote) to predict a continuous outcome variable in situations with zero-inflated predictors and complex correlation structure between predictors.

### 3.2. Data-generating mechanism

Motivated by real-life data from mass-spectrometry studies, we simulated data from a multivariate binomial and lognormal mixture distribution. First, a matrix of latent continuous covariates $Z \in \mathbb{R}^{n \times q}$ with $q = 200$ was sampled from a multivariate log-normal distribution with a prespecified correlation structure (see Supplementary Figure 1), and with a geometric mean of $\exp(10)$ and log standard deviation of 1 for each variable. Omics-derived variables often exhibit clusters of correlated predictors that exhibit stronger dependencies within each cluster than between the clusters. To account for such a group structure, we specified a block-diagonal correlation matrix with a hub correlation structure for 4 groups among the 200 predictors with a group size of 50 predictors, respectively [19]. In each group, the first variable was designated as the hub, while variables 2 through 50 within each group had a progressively declining correlation with the hub variable (see Supplementary Figure 1). We also assumed the presence of one group of with little to no correlation (4[th] group). We defined three settings with the percentage of total zeros (structural and sampling zeros) per predictor within each group ranging from 0 to 25%, 0 to 50% and 0 to 75%, as motivated by the case study (see Supplementary Figure 2). In each scenario, either $1/3$ or $2/3$ of all zeros were structural zeros (see Supplementary Figure 3 and Supplementary Table 1). Structural zero-inflation was accounted for by a $n \times q$-dimensional binary matrix $\boldsymbol{D}$ with elements $d_{ij} \sim Bin(1, p_j^{struc})$, where the parameter $p_j^{struc}$ denotes the proportion of structural zeros in latent variable $j$. The structural zeros were then introduced into the nonnegative, right-skewed intensity values by $\boldsymbol{Z^D} = \boldsymbol{Z} \oplus \boldsymbol{D}$, where $\oplus$ represents the element-wise product of matrices [20].

For the outcome-generating model, we substituted PMVs of $\boldsymbol{Z^D}$ with the theoretical geometric mean of the log-normal distribution, $\mu = \exp(10)$, and then log-transformed $\boldsymbol{Z^D}$, yielding $\boldsymbol{U}$. We assumed that the outcome $\boldsymbol{y}$ for $n$ subjects is generated from the following linear model

$$\boldsymbol{y} = \boldsymbol{\beta_0} + a\boldsymbol{U}\boldsymbol{\beta_U} + (1-a)\boldsymbol{D}\boldsymbol{\beta_D} + \boldsymbol{\epsilon}, \qquad (10)$$

where $\beta_0$ denotes an intercept, $\boldsymbol{\beta_U}$ and $\boldsymbol{\beta_D}$ are two $q$-dimensional vectors of regression coefficients, and $\epsilon \sim N(0, \sigma^2)$ represents an error term. The parameter $a$ controlled the contributions of the continuous and binary components to the variability of the outcome vector $\boldsymbol{y}$. We defined three scenarios based on the value of $a$: (i) the variances of both terms, $a\boldsymbol{U}\boldsymbol{\beta_U}$ and $(1-a)\boldsymbol{D}\boldsymbol{\beta_D}$, were equal ('U=D'), (ii) the variance of $a\boldsymbol{U}\boldsymbol{\beta_U}$ was twice the variance of $(1-a)\boldsymbol{D}\boldsymbol{\beta_D}$ ('U=2D'), or (iii) $a = 1$ ('U') such that only the continuous part was relevant for the outcome generation. In a pilot study, the variance $\sigma^2$ of the error term was determined to approximately achieve a target $R^2$ of 0.3 or 0.6 in a simulated dataset of size $n = 100,000$ for each scenario. Additionally, we considered $\sigma^2 = 0$ to simulate a scenario with $R^2 = 1$ in order to focus on the bias from model misspecification.

Within this general framework, we considered several outcome-generating models (OGM) defining the association between the simulated peptides and the outcome (see also Supplementary Figure 4):

| | |
|---|---|
| OGM (A) | The predictors in groups 1 and 3 were all associated with outcome with varying regression coefficients in [0.1,1] within each group, including both the continuous and binary components. |
| OGM (B) | In each group, only five predictors were associated with the outcome with varying regression coefficients in [1, 2], independently for the continuous and the binary components. |
| OGM (C) | All predictors in groups 1 and 3 were weakly associated with the outcome with varying coefficients in [0.1, 0.4]. The continuous and binary components had identical regression coefficients. |

While the outcome values were generated using model (10) according to the three OGMs, we assumed that a data analyst might also need to deal with sampling zeros in the intensity values. These were generated by setting values which were below the theoretical $p_j^{sam}$-quantile of $\boldsymbol{Z_j}$ to zero. In each scenario, $2/3$ or $1/3$ of all zeros were sampling zeros. Overlap between structural and sampling zeros may occur as structural and sampling zeros were introduced independently on $\boldsymbol{Z}$. We assumed that the hypothetical data analyst in our simulation study received the data matrix $\boldsymbol{Z^a}$ which contained the original right-skewed continuous intensity values $Z$ impeded by (undistinguishable) structural and sampling zeros.

For each combination of the parameters listed above, we considered three sample sizes ($n = 100, 200$ and $400$) but always exactly 200 candidate predictors. We used a full factorial design, resulting in 486 scenarios with 500 repetitions each. The full specification of scenarios is stated in Table 1.

Table 1. Computation and overview of the number of total simulation scenarios

| Source of variability | Number of configurations | Settings |
|---|---|---|
| OGM | 3 | A, B, C |
| Number of candidate predictors | 1 | 200 |
| Maximum proportion of total zero-inflation | 3 | 0.25, 0.5, 0.75 |
| Proportion of sampling and structural zeros | 2 | $(1/3, 2/3)$ and $(2/3, 1/3)$ |
| Specification of the error variance | 3 | $R^2_{target} = \{0.3, 0.6, 1\}$ |
| Specification of $U$ and $D$ contribution | 3 | Only $U$, $U = D$, and $U = 2D$ |
| Sample size | 3 | 100, 200, 400 |
| Total number of simulated scenarios | 486 | |

### 3.3. Estimands

To assess the predictive performance, we considered the root mean squared prediction error (RMSPE), the relative RMSPE, the $R^2$ (i.e. the squared correlation between predicted and observed values) and the calibration slope (CS) of the fitted models on a validation dataset of size $n = 100,000$ that was generated once for each scenario. The relative RMSPE was obtained by dividing the RMSPE of each method by the RMSPE of the benchmark approach oracle-ridge. The CS was computed as the slope of a regression of the observed on the predicted outcome values. To assess predictor selection we recorded the number of selected predictors $q_s(\cdot)$, where a predictor counted as 'selected' if either the continuous or the binary component or both were selected into the final model. We also assessed the true positive discovery rate (TPDR) and the false negative discovery rate (FNDR), which refer to the proportion of selected true predictors among all selected predictors and the proportion of falsely non-selected true predictors among all non-selected predictors, respectively.

### 3.4. Methods

In each of the simulated datasets, the regularized regression approaches defined in 2.2 and 2.3 were evaluated: (i) ridge, (ii) lasso, (iii) lasso-ridge, (iv) ridge-lasso, and (v) ridge-garrote. Prior to model fitting, the generated matrix of predictors $Z^a$ of each simulated data set was transformed in accordance with Section 2.1 to obtain the binary matrix $D$ as well as the two continuous data matrices $X$ and $U$. For $X$, PMVs were substituted prior to the log-transformation by half the global minimum of the non-PMVs in $Z^a$, while for $U$, PMVs were substituted by the predictor-specific geometric means of the non-PMVs in $Z^a$ (see Supplementary Figure 5). Thus, the models containing both $U$ and $D$ considered the 200 predictors with 2 components each, resulting in 400 component variables to be included. The regularization parameter $\lambda$ of the one-stage models was chosen by minimizing the 10-times repeated 10-fold cross-validated RMSPE across a range of 50 potential values for $\lambda$ [21, 22]. For the two-stage approaches, in addition to the selection process used for the one-stage models, we conducted a grid search to identify the optimal combination of $\lambda = (\lambda_1, \lambda_2)$ among $50^2$ possible combinations (see Supplementary Material Section 1.5).

In addition to the regularized regression approaches, we estimated 'oracle models' using ordinary least squares (oracle-OLS) and using ridge regression (oracle-ridge) where exactly the true predictors were included. Table 2 outlines the data utilization within the models, model specifications, regularization parameters, and the coefficients estimated in the final model.

Table 2. Overview of model specifications and data usage at each stage for both the one-stage and two-stage regularized regression approaches

| Approach | Data for the Modelling Stages | | Regularization parameter | Estimated coefficients |
|---|---|---|---|---|
| | Stage 1 | Stage 2 | | |
| Oracle-OLS | True predictors in $U$ and $D$ | - | - | $\boldsymbol{\beta}_U$ and $\boldsymbol{\beta}_D$ |
| Oracle-ridge | True predictors in $U$ and $D$ | - | $\lambda$ | $\boldsymbol{\beta}_U$ and $\boldsymbol{\beta}_D$ |
| Ridge | $U$ and $D$ | - | $\lambda$ | $\boldsymbol{\beta}_U$ and $\boldsymbol{\beta}_D$ |
| Lasso | $X$ | - | $\lambda$ | $\boldsymbol{\beta}_X$ |
| Lasso-ridge | $X$ | Subset of $U$ and $D$ | $(\lambda_1, \lambda_2)$ | $\boldsymbol{\beta}_U$ and $\boldsymbol{\beta}_D$ |
| Ridge-lasso | $U$ and $D$ | $X$ and $D$ | $(\lambda_1, \lambda_2)$ | $\boldsymbol{\beta}_X$ and/or $\boldsymbol{\beta}_D$ |
| Ridge-garrote | $U$ and $D$ | $U$ and $D$ | $(\lambda_1, \lambda_2)$ | $\boldsymbol{\beta}_U$ and $\boldsymbol{\beta}_D$ |

All methods were implemented in the R statistical programming language (R version 4.2.3, R Foundation for Statistical Computing, Vienna, Austria). The code of the simulation study and the implementation of the methods are publicly available as a web supplement on Github: mgregorich/ZIPSel.

## 3.5. Performance measures

As performance measures for the models' predictive capabilities, we computed the mean and standard deviation of the RMSPE, the $R^2$ and the CS across the 500 simulation replicates. The expected values of the TPDR and the FNDR were obtained by taking the mean across simulation replicates in each scenario. Additionally, we determined the predictor inclusion frequency (PIF) [23], representing the probability of each predictor being included, estimated as the proportion of replicates where a predictor was included. Lastly, Monte Carlo Standard Errors (MCSEs) for the RMSPE were computed [18].

## 3.6. Results

Here we present the main findings from the simulation study, for more details we will refer to the supplementary material.

### 3.6.1. Predictive ability

Predictive accuracy evaluated on the validation dataset is illustrated in Figure 1 and Supplementary Figures 6-8. In simulation scenarios with a target $R^2 \in \{0.6, 0.3\}$, the predictive performance of all modelling approaches, except for lasso in the UD dependency setting 'U', was comparable in terms of RMSPE relative to oracle-ridge, as illustrated in Figure 1. For better scale and clarity, the scenarios with a target $R^2 = 1$ have been edited out of Figure 1 but are depicted in Supplemental Figure 6.

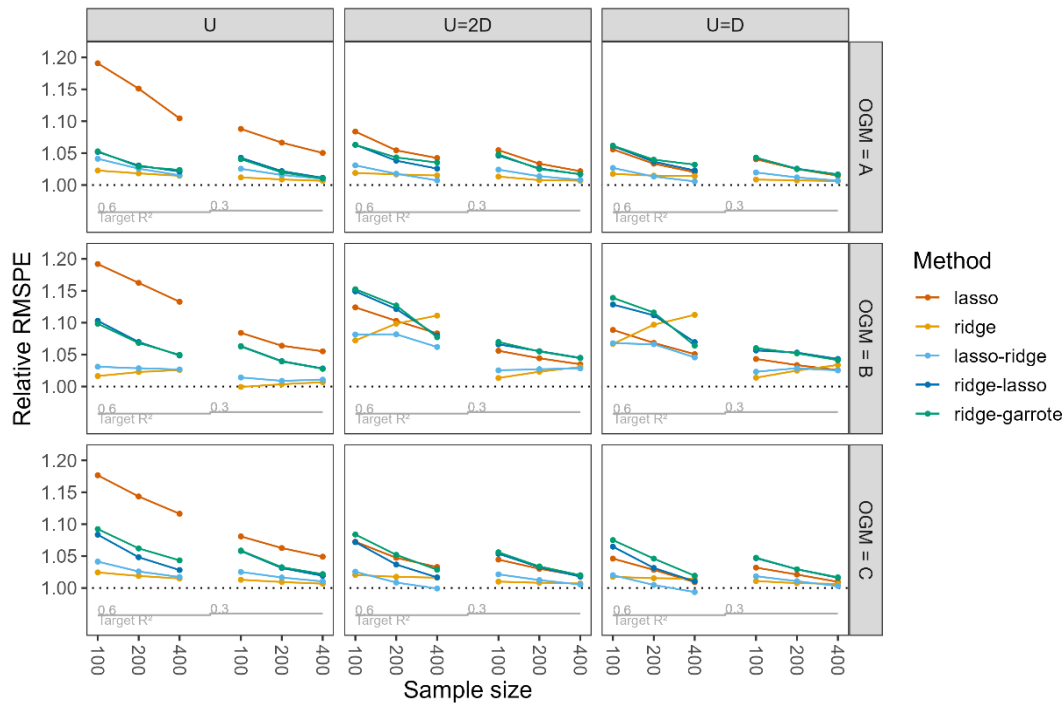

Figure 1. Mean relative root mean squared prediction error (RMSPE) for the three UD dependency settings and three outcome generating mechanisms (OGMs) across sample sizes and target $R^2$ values of 0.6 and 0.3. The maximum percentage of zero-inflation of 75% is fixed of which 1/3 and 2/3 are structural zeros, respectively.

Ridge regression exhibited a slightly superior predictive performance compared to its competitors in OGMs A and C, whereas the lasso had the highest relative RMSPE across all scenarios. Within the two-stage approaches, lasso-ridge outperformed ridge-lasso and ridge-garrote in all scenarios, albeit with only a very small improvement in relative RMSPE. In OGM B, ridge outperformed competitors at low sample sizes ($n = 100$). However, its RMSPE improvements with larger samples were relatively less pronounced compared to those achieved by oracle-ridge (and its competitors), leading to increased relative RMSPE in those cases, as can be seen in Figure 1.

The one-stage methods did not show good calibration across most scenarios, especially noticeable in scenarios with low sample sizes (see Supplementary Figure 7). Specifically, lasso exhibited poor calibration when solely the continuous component was associated with the outcome in the UD dependency scenarios 'U' (see Figure 1 and Supplementary Figure 7). Further, the target $R^2$ values were only approximately achieved by the benchmark models oracle-OLS and oracle-ridge despite increasing sample size, as depicted in Supplementary Figure 8. Even in scenarios without added residual error ($R^2_{target} = 1$), the estimated $R^2$ remained below 1 for all models, due to the information loss induced by sampling zeros and the effects of data transformations.

### 3.6.2. Evaluation of predictor selection

Across scenarios in which both, the continuous component and the binary component were associated with the outcome (UD dependency settings 'U=D' and 'U=2D'), the ridge-garrote approach selected fewer or as many predictors as its competitors, and with the smallest variability, as shown in Figure 2. Lasso-ridge exhibited high variability in the number of selected predictors.

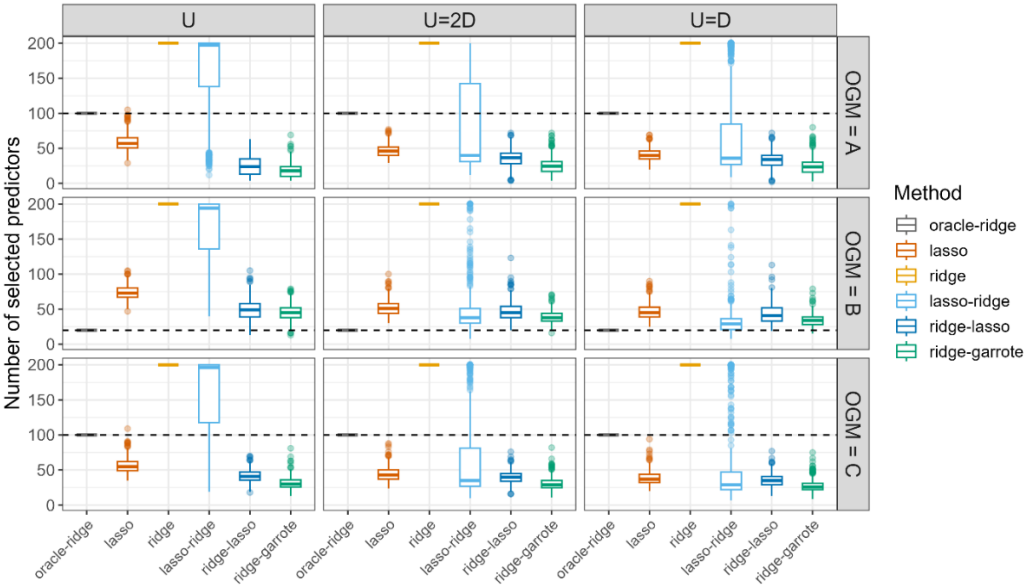

Figure 2. Number of selected predictors across simulation replicates for all one-stage and two-stage modelling approaches for the outcome-generating mechanisms (OGMs) A, B and C and the UD dependency settings 'U', 'U=D' and 'U=2D' pertaining to a setting of $n = 400$, a target $R^2$ of 0.6,

maximum 75% zero values of which 1/3 are structural zero and 2/3 sampling zeros. Black dashed line marks the number of true predictors associated with the outcome for each OGM. Oracle-ridge and ridge are included as benchmark approaches.

While lasso showed low variability in the number of selected predictors, it also tended to have the highest median number of selected predictors followed either by lasso-ridge or ridge-lasso in most scenarios. In most scenarios, ridge-lasso and ridge-garrote had the smallest median numbers of selected predictors. In a direct comparison between the two competitors, ridge-lasso and ridge-garrote, depicted in Supplementary Figure 9, ridge-garrote selected fewer predictors than or as many predictors as ridge-lasso when the continuous and binary component of the predictor were associated with the outcome (setting 'U=D' and 'U=2D'). This was particularly pronounced for higher sample sizes ($n = 400$), while for lower sample sizes ($n = 100$), predictor selection was comparable between ridge-lasso and ridge-garrote.

In Supplementary Figure 10, the PIF of each predictor across simulation replicates for each OGM are illustrated. The PIF for both lasso and lasso-ridge models was influenced by the distribution and magnitude of zero-inflation, while ridge-lasso and ridge-garrote remained largely unaffected. This was evident as the PIF tended to be elevated for predictors exhibiting greater zero-inflation, and conversely, lower for those with lower zero-inflation. This observation was highlighted upon comparing Supplementary Figure 10 with the zero-inflation distribution depicted in Supplementary Figure 2. In terms of TPDR depicted in Supplementary Figure 11, ridge-garrote was more effective than ridge-lasso in selecting true predictors in OGM B when there were few true predictors available. However, the true-positive discovery rate was quite low for all methods in population setting B compared to A and C. In most other scenarios, the TPDR outcomes of both methodologies largely coincided, except in low sample sizes settings in which ridge-lasso demonstrated a higher TPDR.

## 4. Real data analysis: Prediction of kidney function

In this section, we employed the regularized regression approaches and a random forest to predict kidney function based on demographic covariates (age, sex) and a set of predictors obtained through mass-spectrometry. The dataset results from a pooled cohort of urine samples that were analyzed by Mosaiques Diagnostics and comprised 3210 observations on 3192 proteomic predictors. The outcome variable of interest was the estimated glomerular filtration rate (eGFR), a measure of kidney function estimated by the CKD-EPI equation [24], with a mean ± standard deviation in the cohort of $92.50 \pm 23.47$. The set of candidate predictors comprised age ($52 \pm 14$ years), gender (47% female), and the proteomic variables (Supplementary Table 2). Before analysis, we transformed the peptide intensity values as follows (see Supplementary Figure 12): For methods utilizing ***X*** (lasso and stage 1 of lasso-ridge), we log$_2$-transformed the non-PMVs, substituting PMVs with half the global minimum of the

log2-transformed non-PMVs across all peptides. For methods utilizing $U$ and $D$ (ridge, stage 2 of lasso-ridge, ridge-lasso and ridge-garrote), we log$_2$-transformed the non-PMVs for the continuous component, substituting PMVs with the mean of the log2-transformed non-PMVs. Consequently, coefficients linked to the binary component variables $D$ can be understood as the expected difference in the outcome variable when comparing a subject with an average non-zero intensity to one with no intensity for a peptide. We included the demographic variables in the analysis using the 'residual' method as proposed in De Bin et al. [25]. First, a linear model is fitted based on the demographic covariates only, then the resulting linear predictor is included as an offset in a regularized regression model using the peptides as predictors. The continuous variables in $X$ and $U$ were standardized using the standard deviation of the log$_2$-transformed non-PMVs. Predictive accuracy was evaluated by 10-fold cross-validation. For the validation data, the offset of unseen individuals is estimated from the linear model based on the demographic variables and included in the prediction procedure.

In the analysis of mass-spectrometry data peptidomic predictors which exhibit excessive proportions of PMVs are frequently discarded before outcome modelling, as their distribution is nearly degenerate and not informative for prediction. Figure 3 examines the modelling methods across various cutoff levels for the maximum allowed proportion of PMVs (maxPMV) to explore how sensitive results are to these cutoff levels. The highest maxPMV considered was 90%, preselecting 1,333 peptides out of 3,192.

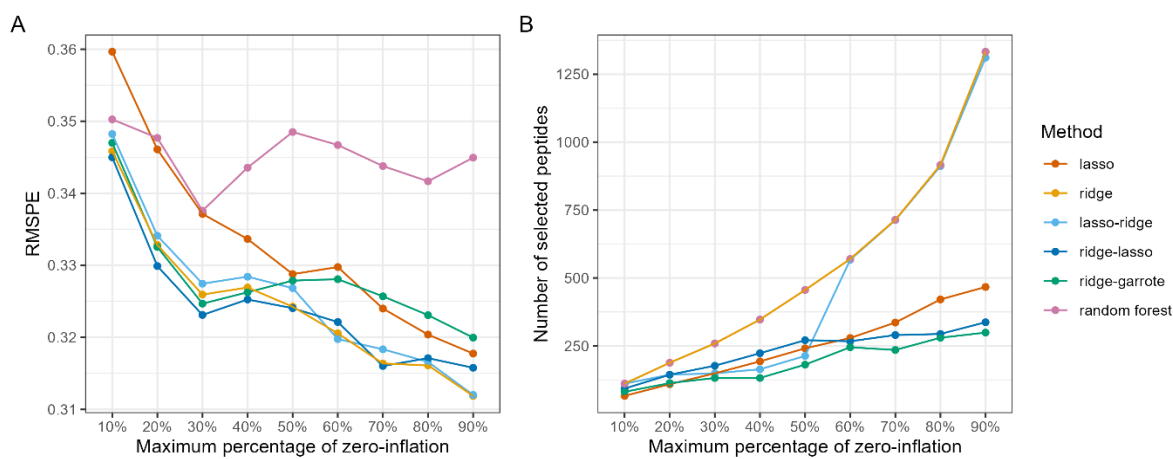

Figure 3. Performance of the regularized regression approaches and the random forest in terms of (A) 10-fold cross-validated root mean squared prediction error (RMSPE) and (B) the number of included/selected peptides (B). The x-axis represents the maximum allowed proportion of PMVs (maxPMV) within a predictor.

The patterns observed in the model evaluation across different maxPMV for lasso, ridge, ridge-lasso, lasso-ridge, and ridge-garrote were similar to the findings of the simulation study. Ridge, lasso-ridge and ridge-lasso mostly overlapped in terms of predictive accuracy and outperformed the competitors (Figure 3, left). Notably, random forest which is not capable of variable selection did not exhibit better predictive performance compared to other methods (Figure 3, left) and did not improve in RMSPE with increasing availability of zero-inflated candidate predictors. Further, the representation of zero-inflated predictors by component variables **U** and **D** (Strategy A of Section 2.1) was always superior to using only **X** (Strategy B) no matter which method was used (see Supplementary Figure 13).

Figure 3 (right) shows that the ridge-garrote model was the approach with the smallest number of selected peptides and demonstrated stability in predictor selection across increasing levels of maxPMV in contrast to lasso-ridge. The lasso-ridge model, despite its predictive capabilities, showed similar variability in terms of the number of selected variables as in the simulation study, as can be seen by the sudden jump in selected predictors at a maximum percentage of zero-inflation of 90%. Furthermore, Supplementary Figure 14 demonstrates that the random forest performed considerably worse in terms of calibration compared to other models. Figure 4 highlights a significant difference in the execution time of the modelling approaches. The fastest of the approaches was random forest, followed by the one-stage methods. Lasso and ridge exhibited notably faster performance compared to the two-stage methods. Among the two-stage modelling approaches, ridge-garrote showed the lowest execution time, while ridge-lasso demonstrated the highest.

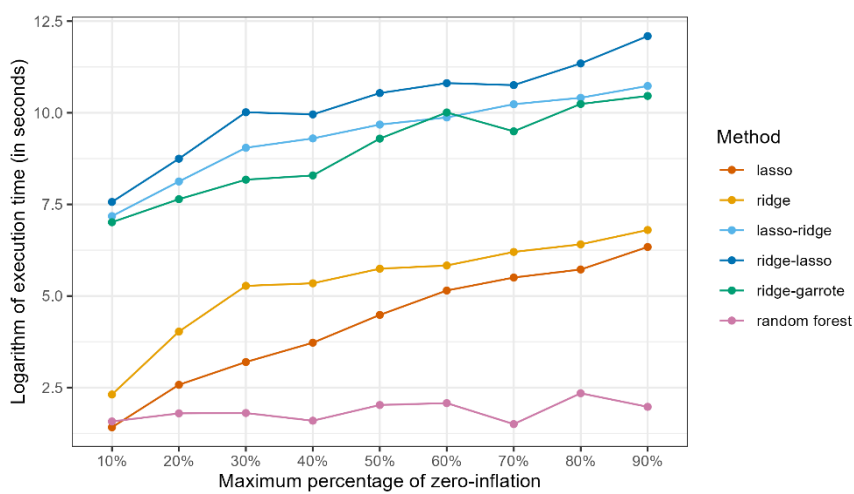

Figure 4. Logarithm of the execution time in seconds of the regularized regression approaches and the random forest. The measured time encompasses the execution of the grid search of the regularization parameter(s) via 10-fold cross-validation, 10 repetitions across a range of 100 potential values for $\lambda$ and the fitting of the final model.

## 5. Discussion

In this study, we established and examined one-stage and two-stage regularized regression approaches, designed to address challenges arising from zero-inflated and correlated predictors. Our emphasis was on evaluating predictive performance and properties of predictor selection. While the findings of the simulation study demonstrated mostly small differences in predictive accuracy and calibration between all methods, the assessment of predictor selection highlighted an advantage of the ridge-lasso and ridge-garrote techniques. However, we noticed that when ridge-lasso had better prediction performance than ridge-garrote this came at the price of selecting more noise predictors into the final model, in particular with higher sample sizes. In the case study regarding the prediction of kidney function based on a high-dimensional set of peptidomic variables, ridge-lasso, ridge and lasso-ridge exhibited the smallest RMSPE followed by ridge-garrote. Despite its good predictive performance, lasso-ridge exhibited large variability in the number of selected predictors both in the simulation study and in the applied case study. Ridge-garrote consistently selected a smaller set of peptides in comparison to its competitors both in the simulation and in the case study. Lasso with imputed PMVs and the random forest showed the poorest predictive performance among the approaches, thus emphasizing the importance of considering the binary component in addition to the continuous intensity values.

A simulation study can only cover a limited number of possible data models. We specified several plausible scenarios regarding the zero-inflation and correlation structure of predictors based on our real data example. The wide variety of scenarios studied likely allows to extend the conclusions on the capabilities of the regularized regression approaches in predictive performance and predictor selection also to data coming from different applications and following different data generating mechanisms.

The appropriate handling of PMVs in predictors continues to be a subject of ongoing research. Single component approaches such as the imputation of PMVs with the sample mean or with $LOD$, $LOD/2$ or $LOD/\sqrt{2}$ has been shown to lead to overestimated standard errors and/or biased estimates for the intercept and the slope in applied studies [4, 13, 26, 27]. By contrast, our study showed that strategies using component variables appeared to be more robust to mixtures of structural and sampling zeros. Recently a latent variable approach was proposed to disentangle such mixtures [13].

The choice of variable selection technique often depends on the specific attributes of the data, the objective of the research, and the balance between model simplicity and predictive accuracy sought by the data analyst. If obtaining a parsimonious model is not a priority, then our simulation study and applied case example revealed that ridge regression with two components representing the continuous and zero-inflated parts of the predictors is perhaps the optimal approach. The utility of ridge-garrote in achieving good predictive accuracy with a minimal set of predictors was apparent in presence of many candidate predictors with varying degree of zero-inflation. If the number of selected predictors is irrelevant, lasso-ridge provided good predictive accuracy. It assumes that all PMVs are sampling zeros at the first stage (which in our simulation study was only partly true), but then relaxes this assumption at the second stage by considering two-component specifications of the selected predictors. A parsimonious selection of a set of predictors that can accurately predict the outcome is often desirable for economic reasons when applying the model, in particular if there is little or no gain when considering all candidate predictors in the model.

In this study we have presented and investigated five regularized regression approaches in their capabilities in dealing with zero-inflated and correlated predictors. While the choice of method may depend on the goal of the analysis to achieve a larger or smaller number of selected predictors, our study demonstrated that optimal predictive accuracy can only be achieved when zero-inflated predictors are appropriately handled using two component variables. Ridge-garrote demonstrated to be an attractive analysis option in situations in which the generation of a shortlist of predictors with good predictive performance is the objective.

**Conflicts of interest**

Harald.Mischak is the co-founder and co-owner of Mosaiques Diagnostics. All other authors have no conflict of interest to disclose.

**Data Availability**

Data of the case study cannot be shared for ethical/privacy reasons. The code of the simulation study is publicly available at GitHub under: mgregorich/ZIPSel,

**Funding**

This work was supported by the Austrian Academy of Sciences from which Mariella Gregorich received funding as part of the DOC fellowship.

# Supplementary Material

## 1. Additional information on the simulation study

### 1.1. The hub correlation structure in the data generation

The hub correlation structure pertained to a setting with 4 groups, where the variables within each group were linked to a central variable (known as the hub) with decreasing levels of correlation strength with increasing distance (in terms of position in the group) to the hub. The hub variable was chosen to be the first variable in each group i.e. the variable with index 1, 51, 101, and 151.

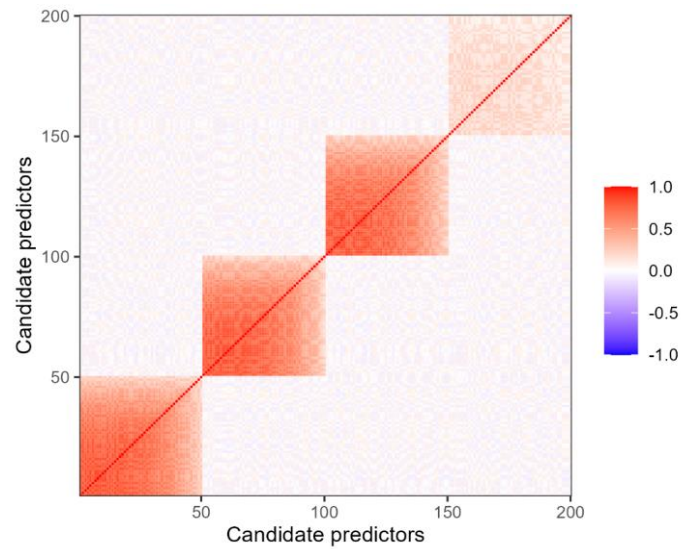

Supplementary Figure 1. Specification of the hub correlation structure of the simulated data with 4 groups.

### 1.2. Specification of zero-inflation of the variables

The shape of the distribution of the proportion of total zero-inflation across candidate predictors within each group was motivated by the observed shape in the dataset from Mosaiques Diagnostics, as illustrated in Supplementary Figure 2. If the proportion of zero-inflation is sorted in an increasing manner and plotted per variable, we can observe an approximately linear increase of zero-inflation in the variables until levelling out at a maximum allowed proportion of zero-inflation of 75%. The maximum proportion of total zero-inflation in the simulation study was restricted to 0.25, 0.5 and 0.75, respectively. Since researchers don't know the split between structural and sampling zeros in a real dataset, the total proportion of zero-inflation was subdivided in varying levels of the proportion of structural and sampling zeros such that one comprised $1/3$ or $2/3$ of the data, while the other made up the rest.

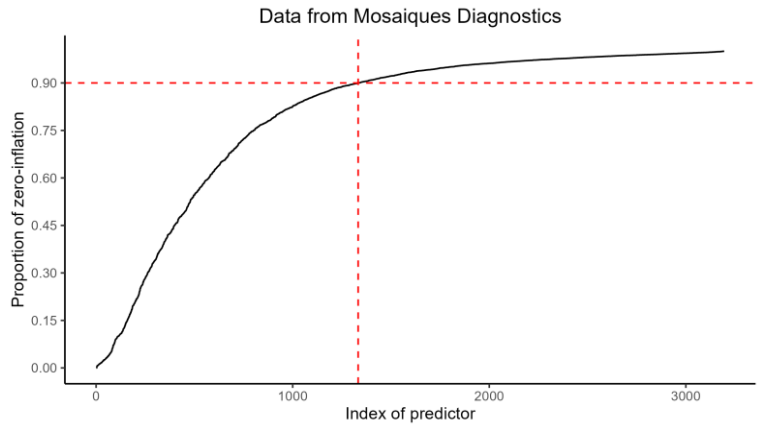

Supplementary Figure 2. Proportion of total zero-inflation observed in the data set from Mosaiques Diagnostics restricted to a maximum allowed percentage of PMVs of 90%. The red dashed line indicates the maximum allowed percentage of PMVs across peptides.

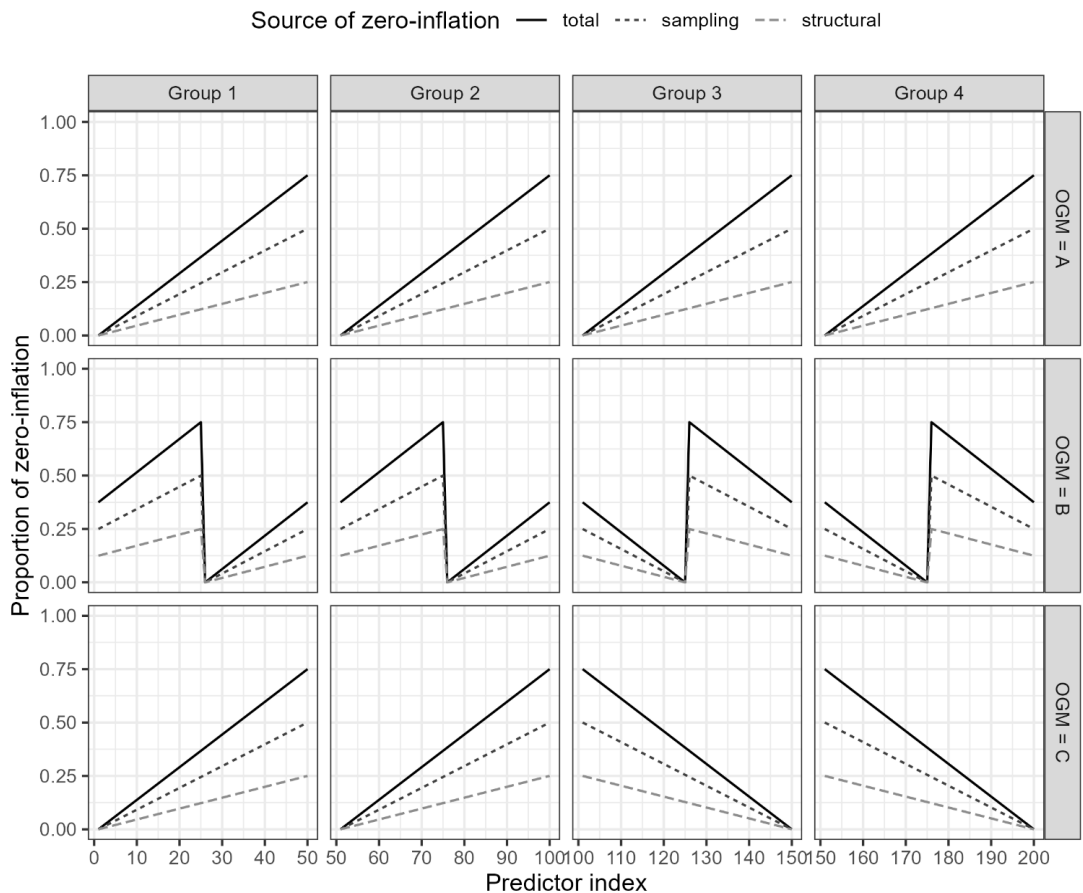

Supplementary Figure 3. Distribution of the proportion of zero-inflation across predictors in each group for a scenario with a structural zero proportion of 1/3 and sampling zeros of 2/3 with a maximum allowed proportion of zero-inflation of 75% across the three outcome generating mechanisms (OGM).

Supplementary Table 1. Overview of zero-inflation specifications with allocation for structural and sampling zeros.

| Maximum percentage of zero-inflation | Structural proportion | Percentage of structural zeros | Sampling proportion | Percentage of sampling zeros |
|---|---|---|---|---|
| **75%** | 1/3 | 25% | 2/3 | 50% |
|  | 2/3 | 50% | 1/3 | 25% |
| **50%** | 1/3 | 16.7% | 2/3 | 33.3% |
|  | 2/3 | 33.3% | 1/3 | 16.7% |
| **25%** | 1/3 | 8.3% | 2/3 | 16.7% |
|  | 2/3 | 16.7% | 1/3 | 8.3 |

## 1.3. Outcome-generating mechanism and variable effect specification: A, B, and C

Different effect specifications of the four predictor groups were established to investigate the regularized regression methods under diverse scenarios, as illustrated in Supplementary Figure 4.

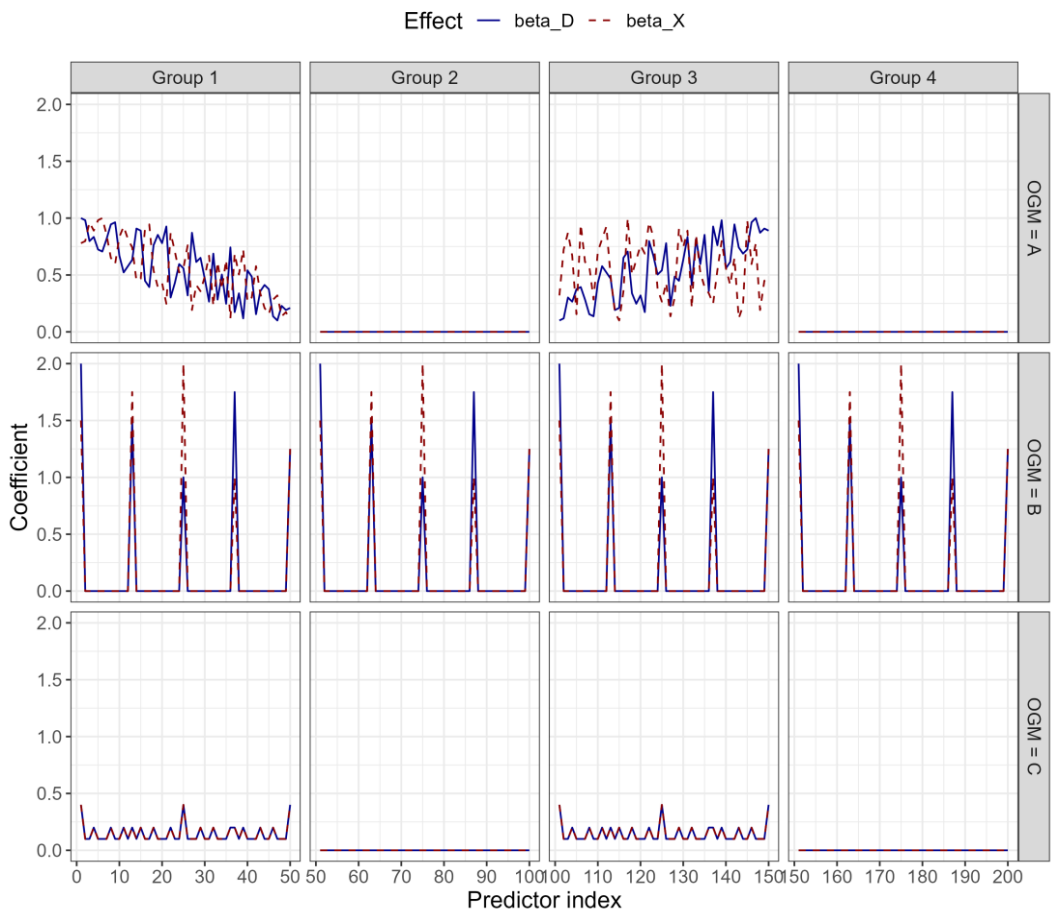

Supplementary Figure 4. True regression coefficients for each OGM setting across predictor groups for the continuous (red) and binary (red) component variables corresponding to each predictor.

## 1.4. Data transformations

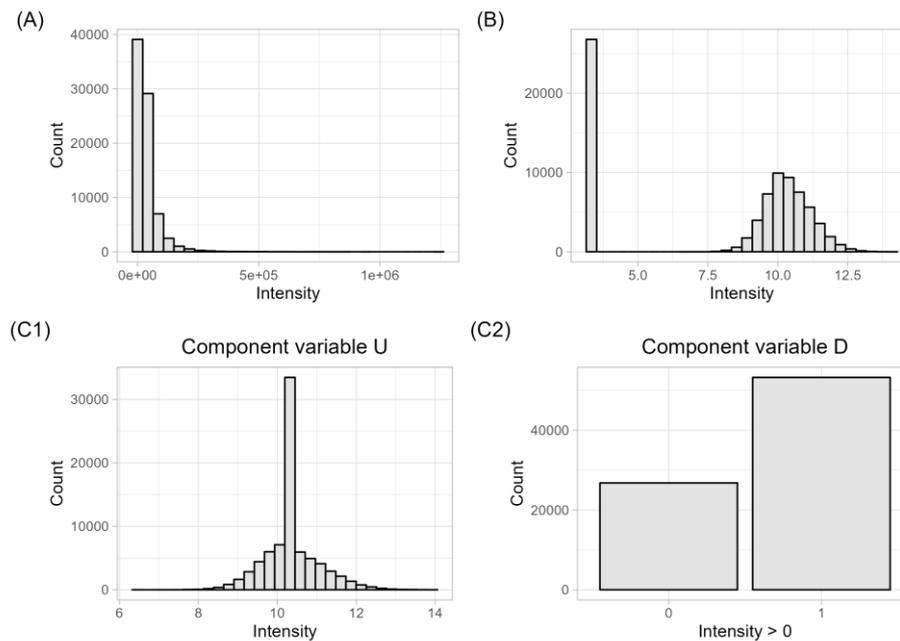

Supplementary Figure 5. Distribution of simulated data with (A) the log-normal distributed data, (B) the log-transformed log-normal distributed data with PMVs at half the global minimum and the two component variables: the continuous component with PMVs at the mean of the log-transformed log-normal data (C1) and the binary component indicating values above 0 (C2).

## 1.5. Penalty parameter tuning for the two-stage approaches

The penalization parameters $\lambda_1$ and $\lambda_2$ can be pre-specified, but are usually optimized simultaneously by minimizing the RMSPE with 10-fold cross validation or the optimization of an information criterion such as Akaike's information criterion. This involves evaluating a series of penalties in the first step (lasso), where each penalty parameter is assessed against a sequence of penalties in the second step (ridge). This process generates a matrix of potential penalty parameters. The optimal combination is determined by iteratively calculating the cross-validated root mean squared prediction error (RMSPE) for each pairing, with the process being repeated R times. The RMSPEs across each iteration and repeated run for each combination $(\lambda_1, \lambda_2)$ are then averaged and the pair resulting in the minimum RMSPE is selected.

## 1.6. Results of the simulation study

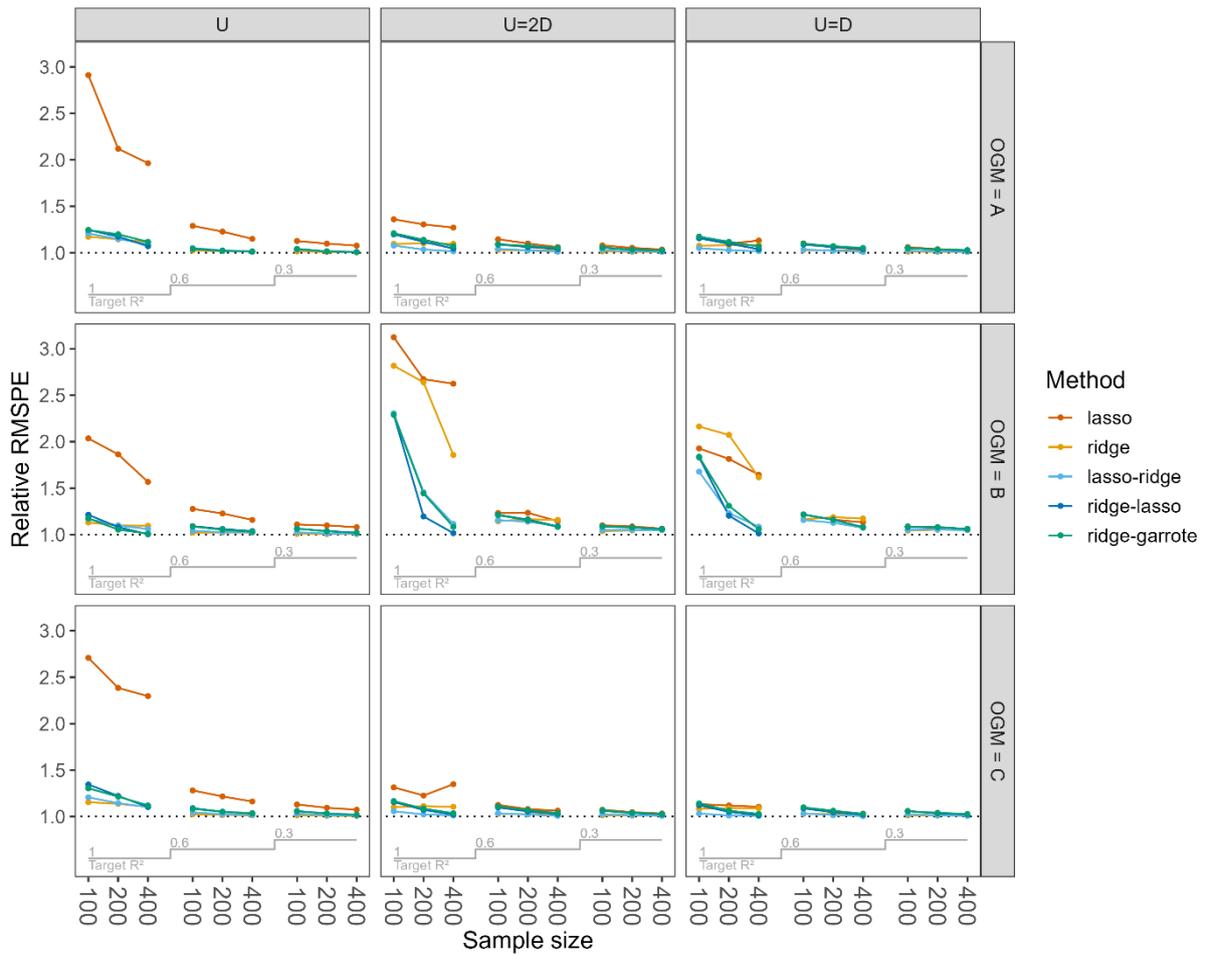

Supplementary Figure 6. Relative RMSPE for all methods. Relative RMSPE was obtained by dividing the method's RMSPE by that of the benchmark method ridge-oracle in each simulation replicate on the validation data ($n = 100,000$). Results are given for different outcome generating mechanisms (OGM A, B, C) and UD dependency setting ('U', 'U=D', 'U=2D') with a fixed maximum proportion of zero-inflation of 75% of which 2/3 comprise structural zeros and 1/3 sampling zeros.

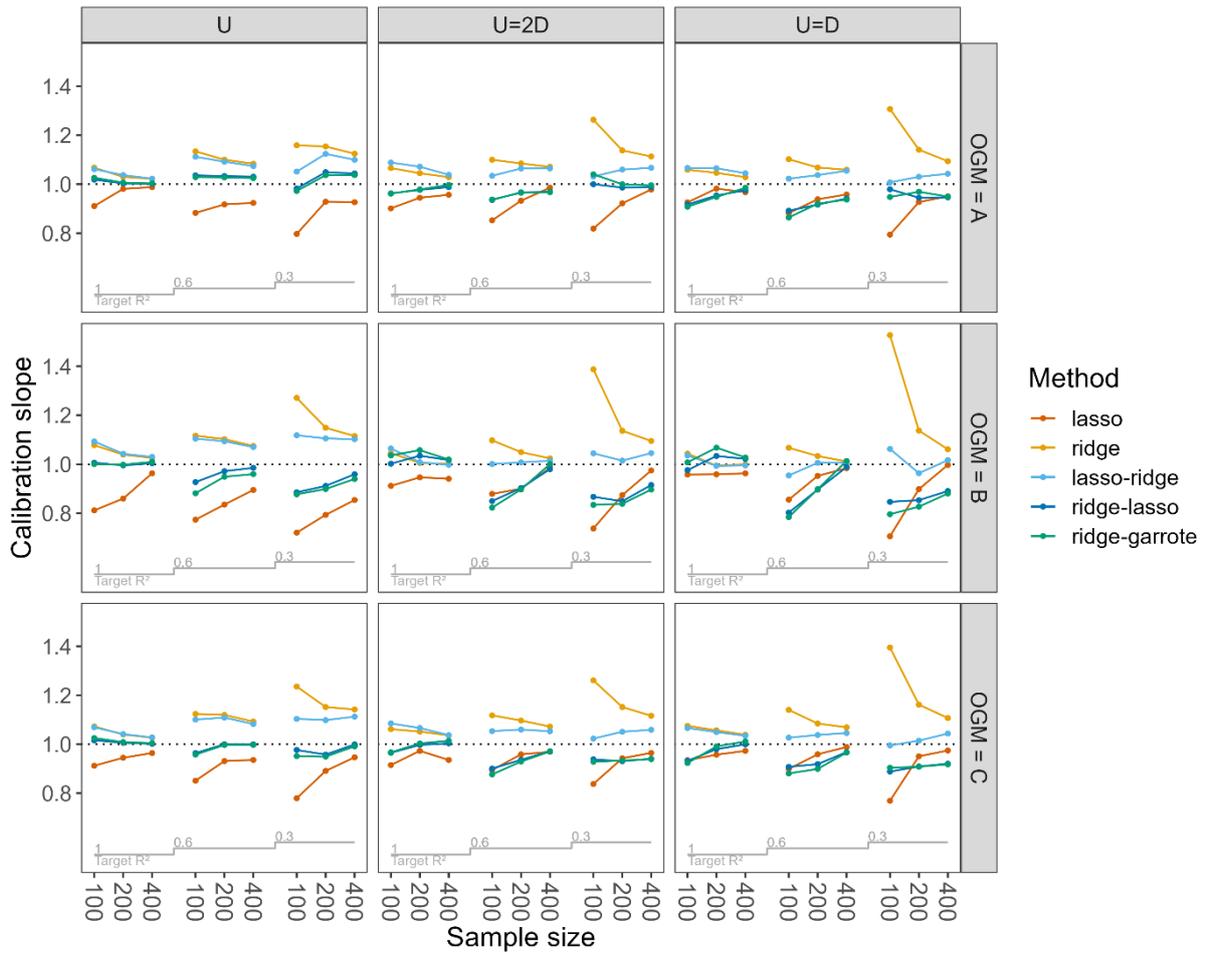

Supplementary Figure 7. Mean calibration slope for all methods evaluated on the validation data ($n = 100,000$). Results are given for different outcome generating mechanisms (OGM A, B, C) and UD dependency setting ('U', 'U=D', 'U=2D') with a fixed maximum proportion of zero-inflation of 75% of which 2/3 comprise structural zeros and 1/3 sampling zeros.

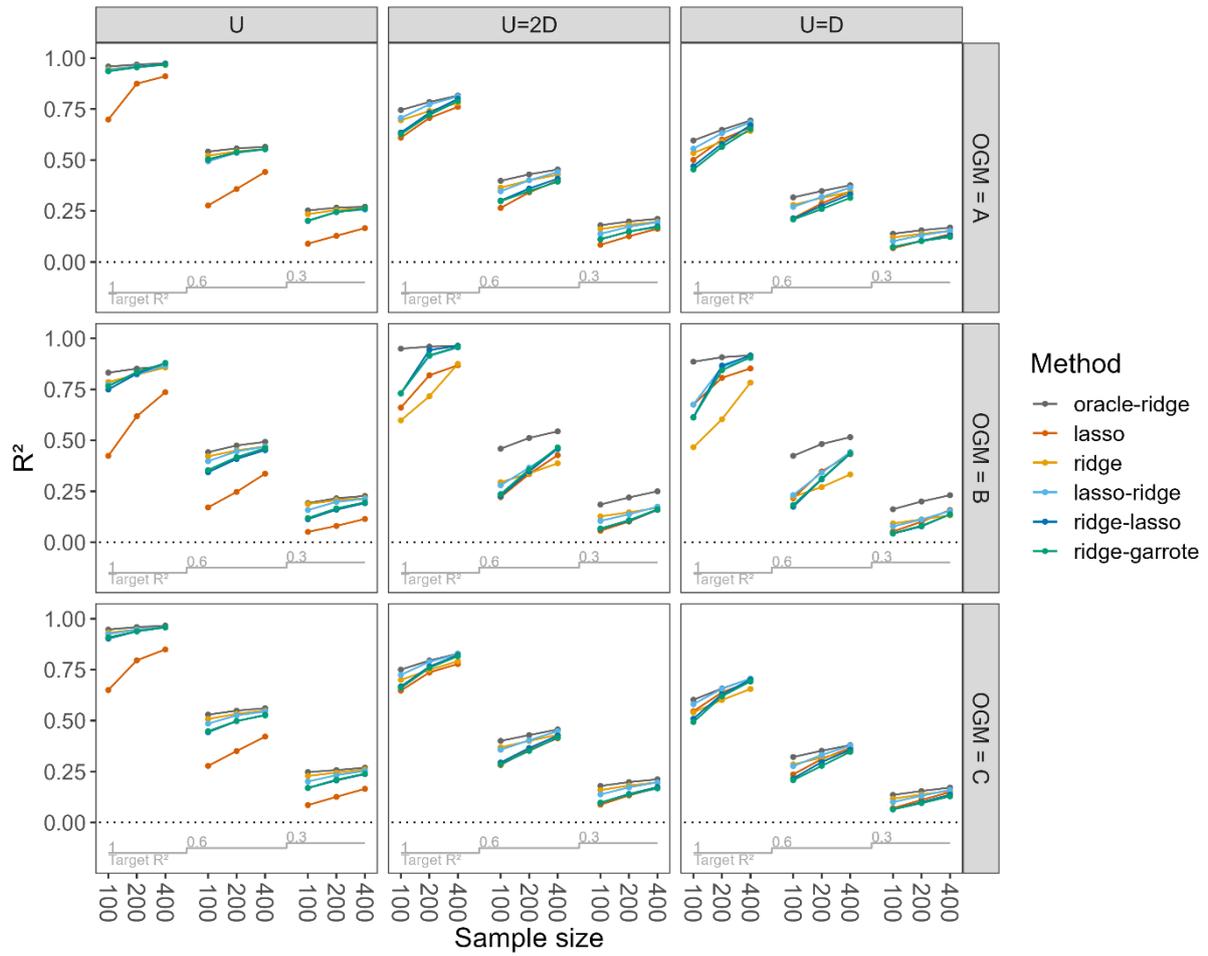

Supplementary Figure 8. Achieved $R^2$ for all methods on the validation data ($n = 100,000$). R² was defined as defined as the squared correlation in predicted and observed values. Results are given for different outcome generating mechanisms (OGM A, B, C) and UD dependency setting ('U', 'U=D', 'U=2D') with a fixed maximum proportion of zero-inflation of 75% of which 2/3 comprise structural zeros and 1/3 sampling zeros.

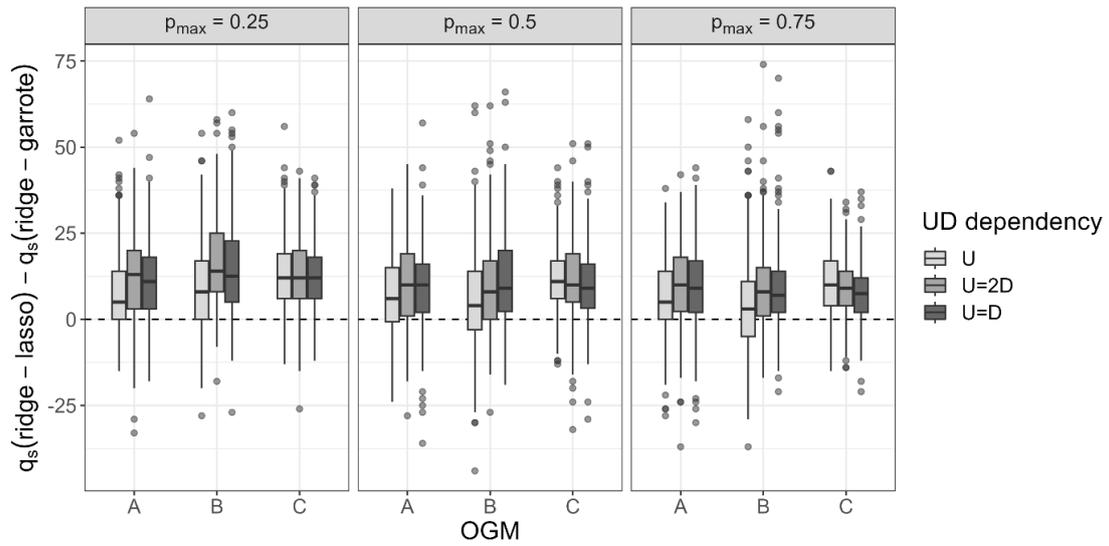

Supplementary Figure 9. Direct comparison in the number of selected peptides between ridge-lasso $q_s(ridge\text{-}lasso)$ and ridge-garrote $q_s(ridge\text{-}garrote)$ for $n = 400$, fixed proportion of structural zeros of 1/3 of the maximum proportion of zero-inflation and no added residual error. A positive difference indicates the selection of more predictors for the ridge-lasso approach, while negative values indicate more peptides selected in the ridge-garrote approach.

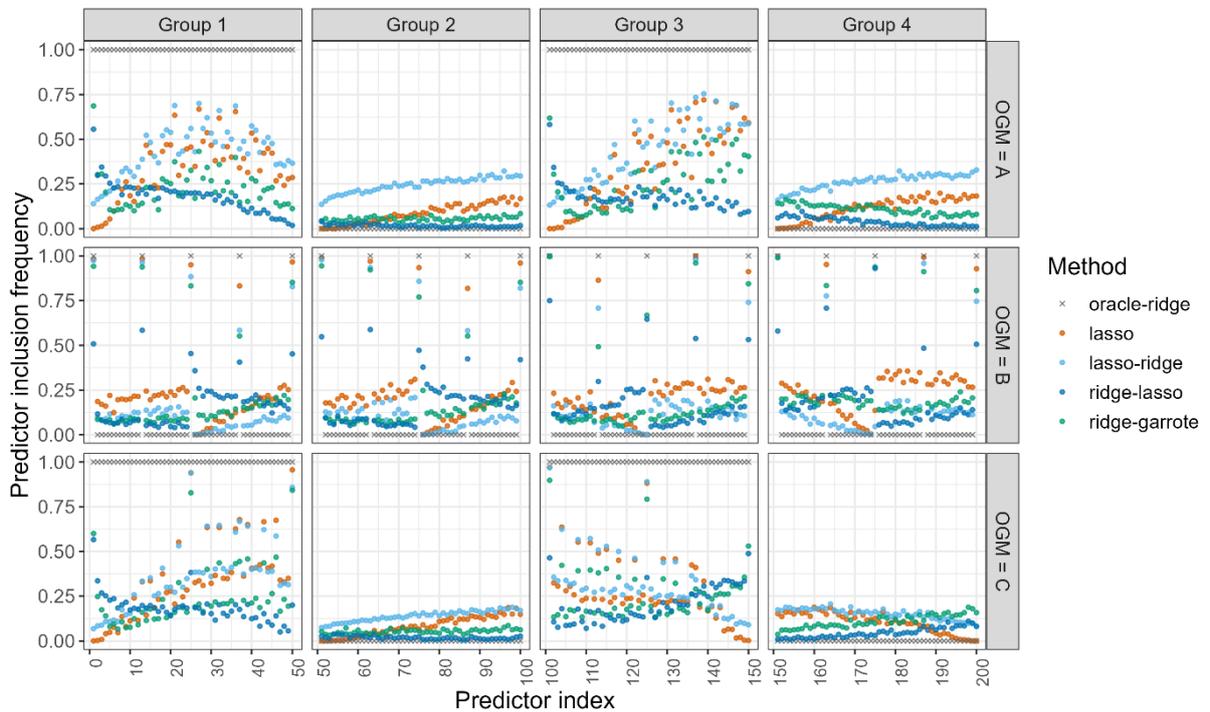

Supplementary Figure 10. Predictor inclusion frequency across simulation replicates in all outcome generating mechanisms A-C for each of the methods performing predictor selection with a fixed sample size of $n = 200$, a maximum proportion of zero-inflation of 75% with 1/3 structural zero proportion, UD dependency 'U=D' and target R2=1. Predictor inclusion frequency of oracle-ridge was included in the figure in grey to indicate the true predictors across OGMs and groups.

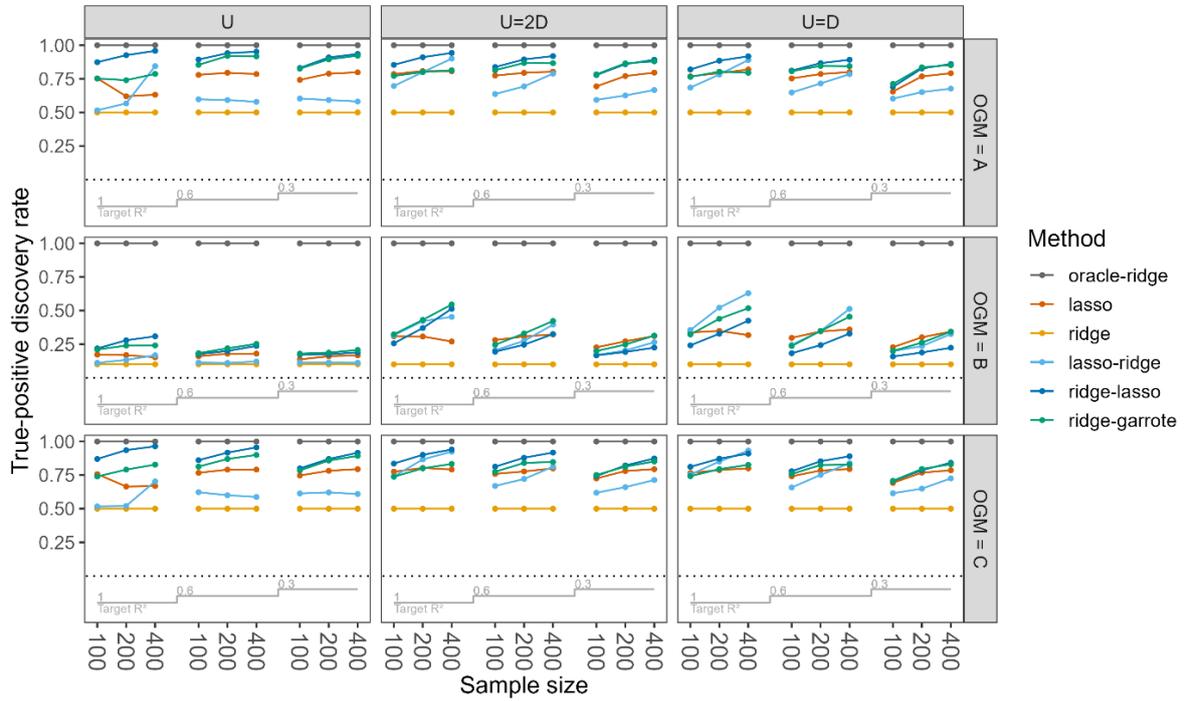

Supplementary Figure 11. True-positive discovery rate of all regularized regression methods across simulation replicates for all outcome generating mechanisms (OGM A, B, C) and for a setting with maximum proportion of zero-inflation of 75% of which 1/3 are structural zeros and 2/3 sampling zeros.

## 2. Additional information on the real-life data example: Prediction of kidney function

### 2.1. Data and modelling strategy

The dataset contains data from 3,210 individuals, encompassing measurements of 3,192 peptides obtained through mass spectrometry.

Supplementary Table 2. Overview of patients' characteristics of the data from Mosaiques Diagnostics

| Variable | Value |
| --- | --- |
| Sample size (n) | 3,210 |
| Peptides (n) | 3,192 |
| Age, years | 52.03 ± 13.46 |
| Sex, female (%) | 1,510 (47.0) |
| eGFR, 1.73m2/min/l | 92.50 ± 23.47 |

Out of the total 3,192 peptides, only 1,333 exhibited more than 10% non-zero values (i.e., intensity values above 0) and were incorporated into the five regularized regression models and the random forest.

Due to the skewed nature of the original peptides' intensity values **Z**, we used a $\log_2$ transformation for the continuous component **U** of the peptide intensities defined as

$$u_{ij} := \begin{cases} \log(z_{ij}), & z_{ij} > 0, \\ m(z_j^+), & z_{ij} = 0, \end{cases} \quad (1)$$

where

$$m(z_j^+) := \frac{1}{n^+} \sum_{i=1}^{n} \log_2(z_{ij}) I(z_{ij} > 0) \quad (2)$$

with $n^+$ indicating the number of intensities $z_{ij}$ larger than 0. This choice of replacement of zero signals in $\boldsymbol{u_j}$ mainly aids the interpretability of the regression coefficients, but also other definitions are possible (e.g. the median of log-transformed positive intensities). For a model including **U** and **D**, the coefficients corresponding to the binary variable can be interpreted as the expected difference in the outcome variable when comparing a subject with an average non-zero signal to a subject with absent signal for a peptide. For **X**, data was $\log_2$-transformed and PMVs were substituted with half the global minimum of the $\log_2$-transformed non-PMVs.

Supplementary Figure 12 illustrates the impact of the data transformations on the distribution of a peptide's original intensity values displayed in Panel A. Panel B exhibits Strategy B resulting in a more normal distribution for positive values, with zeros replaced by half of the minimum $\log_2$-transformed value. Panels C1 and C2 depict the Strategy A with the continuous and binary component variables, respectively.

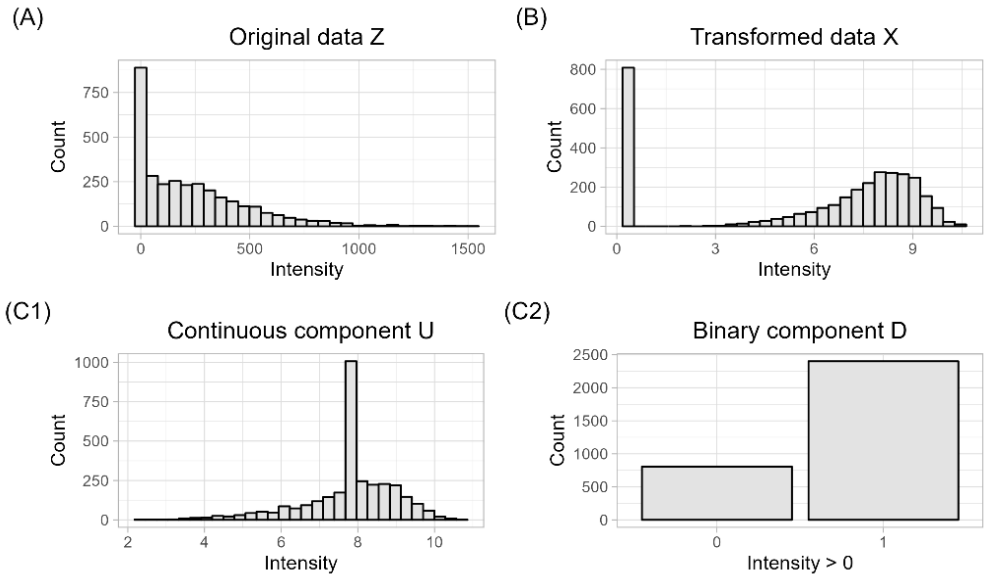

Supplementary Figure 12. Illustration of the original intensity values of one selected peptide exhibiting 25% PMVs and the implications of the strategies of handling zero-inflated predictors in the dataset provided by Mosaiques Diagnostics,.

We used the random forest to investigate whether an ensemble method not requiring data pre-processing could yield improved prediction performance. The number of trees was set to 1,000 as a compromise between performance, stable variable importance estimates and computation time [17, 18]. Bootstrap samples were drawn with replacement. We expected only minor improvements by optimizing the values of the hyperparameters for random forest according to [19]. Thus, only the default setting for mtry as $\left\lfloor \frac{p}{3} \right\rfloor$, a node size of 5 and a splitting rule minimizing the variance was implemented.

## 2.2. Results of the data example

The real-life data example demonstrated the importance of strategy A to represent the distribution of a zero-inflated predictor by two component variables. We did not only fit the models according to the component variable strategy but also implemented all methods using the $\log_2$-transformed data, with PMVs substituted by half of the minimum value to assess the impact of the component split.

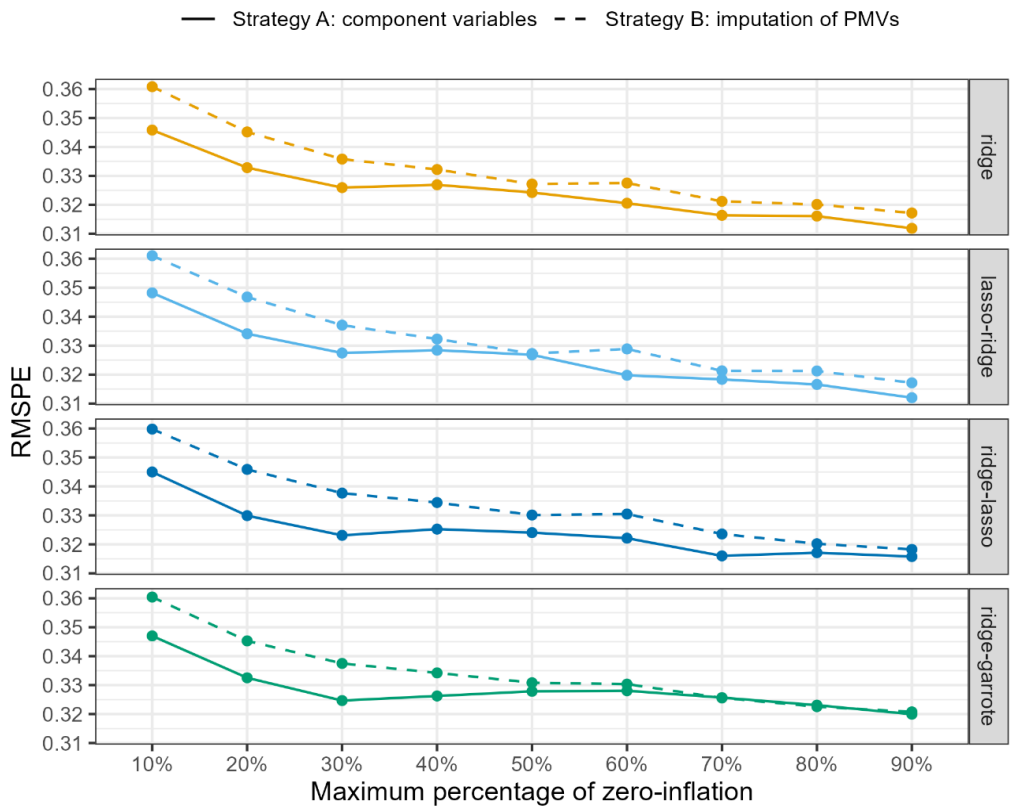

Supplementary Figure 13. RMSPE for all modelling approaches and random forest is depicted, with solid lines representing approaches using the component split strategy, and dashed lines denoting the imputation approach for PMVs.

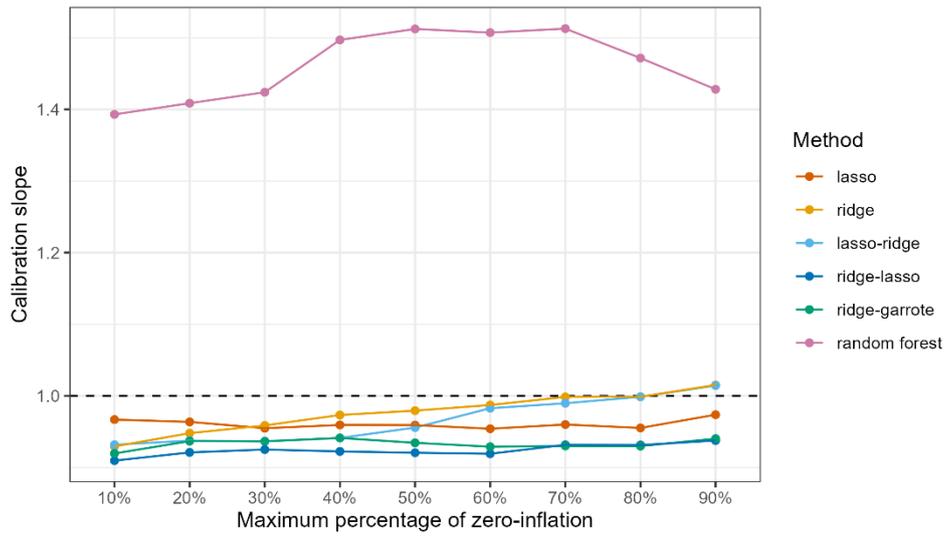

Supplementary Figure 14. Calibration of the modelling approaches and random forest in terms of